\documentclass[manuscript]{copernicus}  

\nolinenumbers
\usepackage{color}
\usepackage[T1]{fontenc}

\begin{document}

\title{On the applicability of Taylor's hypothesis in streaming magnetohydrodynamic turbulence
}

\author[1,3]{R. A. Treumann}
\author[2]{Wolfgang Baumjohann}
\author[2]{Yasuhito Narita}
\affil[1]{International Space Science Institute, Bern, Switzerland}
\affil[2]{Space Research Institute, Austrian Academy of Sciences, Graz, Austria}
\affil[3]{Geophysics Department, Ludwig-Maximilians-University Munich, Germany\\

\emph{Correspondence to}: Wolfgang.Baumjohann@oeaw.ac.at
}

\runningtitle{Taylor hypothesis}

\runningauthor{R. A. Treumann, W. Baumjohann, Y. Narita}

\received{ }
\pubdiscuss{ } 
\revised{ }
\accepted{ }
\published{ }


\firstpage{1}

\maketitle

  
\noindent\textbf{Abstract}.-- 
We examine the range of applicability of Taylor's hypothesis used in observations of magnetic turbulence in the solar wind. We do not refer to turbulence theory. We simply ask whether in a turbulent magnetohydrodynamic flow the observed magnetic frequency spectrum can be interpreted as mapping of the wavenumber turbulence into the stationary spacecraft frame. In addition to the known restrictions on the angle of propagation with respect to the fluctuation spectrum and the question on the wavenumber dependence of the frequency in turbulence which we briefly review, we show that another restriction concerns the inclusion or exclusion of turbulent fluctuations in the velocity field. Taylor's hypothesis in application to magnetic (MHD) turbulence encounters  its strongest barriers here. It is applicable to magnetic turbulence only when the turbulent velocity fluctuations can practically be completely neglected against the bulk flow speed. For low flow speeds the transformation becomes rather involved. This account makes even no use of the additional scale dependence of the turbulent frequency, viz. the existence of a ``turbulent dispersion relation''.  

\section{Introduction}
{\citet{taylor1938},  struggling with stationary turbulence, suggested that its wavenumber spectrum could be directly inferred, if only the turbulence were embedded into a sufficiently fast bulk flow.} 

{Stationarity implies that the total time derivative vanishes. With $\vec{V}=\vec{V}_0+\vec{U}_0+\delta\vec{V}$ the total velocity,  $\vec{V}_0$ bulk and $\vec{U}_0$ mean large-scale (turbulent mechanical energy carrying) eddy velocities, one trivially has 
\begin{equation}
\frac{d\delta\vec{V}}{dt}=\frac{\partial\delta\vec{V}}{\partial t}+\vec{V}\cdot\nabla\delta\vec{V}=0
\end{equation}
for the turbulent fluctuations  $\delta\vec{V}$ of $\vec{V}$. Neglecting the nonlinearity in the convective term then, for any observer at rest at location $\vec{x}_s=\vec{x}\pm\vec{V}t$, the flow maps the original turbulent wavenumber $\vec{k}$-spectrum onto an easily detectable stationary observer's (spacecraft) frequency $\omega_s$-spectrum
\begin{eqnarray}\label{eq-tay-1}
\delta{V}(\vec{x}_s,t)&=&\frac{1}{(2\pi)^4}\int d\omega_k\,d\vec{k}\exp\Big[-i\omega_kt+i\vec{k\cdot x}\Big]\nonumber\\[-2ex]
&&\\[-2ex]
\omega_s&=&\omega_k\pm \vec{k\cdot V}\nonumber
\end{eqnarray}
where $\omega_k$ is the possible internal frequency of turbulence, the turbulent ``dispersion relation''.\footnote{{The notion of a turbulent dispersion relation seems to be alien to turbulence theory which just considers a wavenumber spectrum which is understood as the integral $\int d\omega (\dots)$ with respect to frequency of the power spectral density of the turbulent fluctuations. In $\omega,\vec{k}$ space the power spectral density occupies a volume which, resolved for $\omega(\vec{k})$ gives a complicated multiply connected relation between frequency and wavenumber, the turbulent dispersion relation. This is not a solution of a linear wave eigenmode equation whose solutions are ordinary waves.}} Assuming that the latter is negligible $\omega_k\ll |\vec{k\cdot V}|$ compared with the total speed of the flow, observation of the frequency spectrum then apparently directly reproduces the wavenumber spectrum of velocity turbulence.}

{In spite of its appealing simplicity and its just mentioned logical assumptions, Taylor's hypothesis (as it is commonly called) has a number of critical implications when applied to non-mechanical fluctuations and turbulence like magnetic power spectral densities.  These are frequently  overlooked or entirely ignored, thus making sense to check for the validity and applicability of Taylor's hypothesis in these cases. These implications concern the following:}

{1. Though probably not the most important in fast flows, Taylor's hypothesis applies in this form only to time intervals when the turbulence can indeed be considered stationary. }

{2. It ignores any nonlinearity in the stationarity condition, which can, however, be justified again as a reasonable approximation to sufficiently fast flows and sufficiently small fluctuation amplitudes. }

{3. More crucially, the exponential in the Fourier representation of the velocity fluctuations depends itself on the velocity fluctuations, as is obvious from the Galilei transformed observer's frequency. This means that it contains self-interactions of the fluctuations. These, for the turbulent velocity field can be neglected or summed up into the large-scale energy-carrying mean eddy velocity $\vec{U}_0$ \citep{tennekes1975}. Its neglect in this place for the magnetic fluctuations $\delta\vec{B}$ must, however, be questioned for two reasons: It sensitively affects the fluctuation phase, and in addition, it causes correlation between magnetic and velocity fluctuations because of their different scales. Consistency requires that the argument of the exponential in the magnetic fluctuation field should be expanded up to second order in $\delta\vec{V}$. We will account for this effect below. }

{4. The transformation is, within these limitations,  justified well for the turbulent velocity fluctuations $\delta\vec{V}$. Through the continuity equation, it can also be justified for the fluctuations of density $\delta N$ in compressible turbulence and through that, under additional assumptions on the equation of state, also for the fluctuations $\delta T$ of temperature.} 

{5. Any straightforward application to $\delta\vec{B}$, the turbulent magnetic field fluctuations \citep[see, e.g.,][]{roberts2014}  can be defended only under rather severe restrictions, as will be demonstrated below. What concerns this last point, it requires some explanation. In application to the electromagnetic field we note the following:}

{(a) It is common knowledge that the electromagnetic field in moving media is not Galilei invariant. It is Lorentz invariant for all flows, whether relativistic or nonrelativistic. } 
 
{ (b) This general remark on invariance seems to disqualify  \citep[though maybe not expressed as rude as by][]{saint2000} any application of Taylor's hypothesis in its nonrelativistic version to electromagnetic turbulence\footnote{Doubts in the \emph{general} validity of Taylor's hypothesis and its unreflected application have been expressed not only for MHD  \citep[e.g., by][]{goldstein1986,lvov1999,nariyuki2006,matthaeus2010,wilczek2012,wilczek2014,klein2014,klein2015,huang2015,lugones2016,narita2017,narita2018,perri2017,treumann2017} but also in other fields \citep[][]{belmonte2000,saint2000,tsinober2001,burghelea2005,dennis2008,he2010,davoust2011,macmahan2012,higgins2012,geng2015,creutin2015,goto2016,podesta2017,shet2017,squire2017,cheng2017,bourouaine2018,yang2018,kumar2018} dealing with turbulence, among them meteorology, hydrology, channel flows, river research and others.}  and in particular\footnote{See also the thermodynamic arguments in \citet{treumann2017} referring to observed spectra \citep{safrankova2016}.}  to magnetohydrodynamic (MHD) turbulence. However, the correct relativistic approach in \emph{ideal} MHD just introduces the induction electric field, which can be considered an additional constraint to be satisfied \emph{afterwards} when accounting for the turbulent spectrum of the electric field. This allows the separate consideration of $\delta\vec{B}$ as functional of $\delta\vec{V}$ in ideal MHD.}
 
{(c) The electromagnetic field in classical physics does never become turbulent by itself. Without any exception, the electromagnetic field is secondary to turbulence, being the consequence of formation of turbulent vortices, electric currents, and possibly even weak large-scale charge-separation fields in a conducting turbulent medium (plasma, conducting fluid $\dots$) as consequence of the turbulent velocity field, as well as density and temperature gradients, i.e. inhomogeneities. Turbulence is basically mechanical. The electromagnetic field reacts passively to it. Transformation of the turbulent velocity field into the observer's (spacecraft) frame according to Taylor-Galilei can, within weak assumptions, stand up. Its effect on the electromagnetic field is by no means straightforward to account by transformation from the turbulent source into a moving frame.}

{(d) The reason for applying Taylor's hypothesis to magnetic fluctuations in MHD turbulence is justified by the wish to infer the otherwise difficult to access power density spectrum as function of wavenumber $\vec{k}$ by interpreting the frequency spectrum as the Galilei transformed wavenumber spectrum.} 

{In the following we discuss the validity of Taylor's hypothesis for the turbulent fluctuation of velocity $\delta\vec{V}$ and magnetic field $\delta\vec{B}$. We show that for the former its application is justified, while for the latter its application is subject to severe restrictions.} 

{We do \emph{not} refer to any turbulence theory nor model equations except when taking them as an input.  More is not required for the limited purpose of this note. It would introduce further unnecessary complications. Taylor's hypothesis is not intrinsic to turbulence theory. Its validity (or invalidity) can be demonstrated independently of any theory of turbulence when occupying an observer's point of view, asking the question, in which way turbulent fluctuations do transform into the observer's frame such that from observation of fluctuations in frequency space one may infer the turbulent wavenumber spectrum.}

{This is a simple practical question. What is meant here is, in the first place, the spectrum of fluctuations. It is  \emph{not} the power-density  spectrum. In order to, in a second step, calculate the power density spectrum, which is central to turbulence theory, one must know how the fluctuations themselves transform.} 

{After having clarified this question for the velocity fluctuations, we then ask for their effect on any related magnetic field fluctuations. We show that the modification of the magnetic fluctuation spectrum by application of the Taylor hypothesis runs into  complications. Only under strict and severe assumptions, Taylor-Galilei transformation of the magnetic field makes some approximate sense but requires caution in the interpretation of the results.}

\section{A brief review of Taylor's proposal}

{In recent years sophisticated measurements of solar wind turbulence \citep[cf., e.g.,][for reviews]{goldstein1995,tu1995,zhou2004,podesta2011} have advanced our knowledge about the evolution of turbulence in a highly conducting (i.e. collisionless) magnetised ideal plasma, in this case the fast streaming solar wind as a paradigm of fast streaming magnetised stellar winds. Since the latter are barely accessible (and probably, on the time scale of observations and spatial scale of expansion, by no means ideally conducting and collisionless), such measurements are  important for understanding their dynamics, evolution of turbulence, its contribution to dissipation, entropy generation, and possibly even generation of observable thermal and nonthermal radiation. In other fields like fluid mechanics, hydrology, and meteorology, which are all dealing with turbulence, information obtained comparably easily from solar wind turbulence is as well valuable.}

{The fast solar wind stream seen by stationary observers (spacecraft)  on multi-scales \citep{schwartz2009} is advantageous when transporting the frozen-in turbulence across the fixed frame. What is usually observed, are temporal fluctuations which can be transformed into frequency spectra. Taylor's hypothesis  \citet{taylor1938} comes in here for help \citep{roberts2014} when one wishes to infer the spectrum in wavenumber space.} 

{This problem has become of interest since roughly two decades in coincidence with spectral observations of solar wind turbulence reaching down into the assumed dissipative range  \citep[cf., e.g.,][]{alexandrova2009,sahraoui2009,sahraoui2013,huang2015}, multi-spacecraft observations were combined to directly measure  spatial spectra, observations of turbulent electric fields \citep{chen2011} became available, and turbulence spectra in both the velocity \citep{podesta2007,podesta2009} and plasma density \citep[first \emph{in situ} observations of electron density spectra, already exhibiting all much later confirmed details,  date back to][]{celnikier1983} were measured directly \citep{chen2012,safrankova2013,safrankova2016}.} 

{According to Taylor's hypothesis the frequency of change in the velocity fluctuations measured in the spacecraft frame is given by Eq. (\ref{eq-tay-1}). 
It holds reasonably well if either $\omega_k$ is known, or otherwise the internal turbulent variations are not remarkable $\omega_k\ll\vec{k\cdot V}$ compared with the flow. In the dissipative regime this will not be true anymore. Molecular scales are of no interest here, but dissipation sets on at much longer scales in the Hall regime already  \citep{alexandrova2009,narita2006,sahraoui2012} and on the presumable electron gyro-scales \citep{sahraoui2009,sahraoui2013} where anomalous dissipation takes over as the ultimate sink of turbulent \emph{magnetic} energy \citep[for a particular argument, see,][]{treumann2017}, which is just a fraction of the mechanical energy stored in the turbulence. }

The observation that the change in frequency depends on the angle between the turbulent wavenumber and the streaming velocity does not substantially violate Taylor's assumption; it, however,  affects the isotropy of the observed turbulence. The angular dependence of Taylor's hypothesis tells the trivial truth that any stationary turbulent eddies which propagate at angles larger than 
\begin{equation}
\theta_V> \cos^{-1}(\omega_k/kV) =\cos^{-1}(1-\omega_s/kV)
\end{equation}
remain unaffected by the transformation into the spacecraft system, a boundary which can be easily obtained from observations if distinguishing angles, thus limiting the reliable wavenumber range. All turbulent fluctuations near this angular boundary become mapped to either zero frequency $\omega_s\approx0$ or $\omega_s\approx2\omega_{{k}}$, depending on the direction of wave number with respect to the streaming velocity, for $\omega_k\neq0$ causing a frequency dependent deformation and directional anisotropy of the spectrum of turbulent fluctuations, which to distinguish from other anisotropies poses an own problem.

For a system of eddies each of which rotates at some mean eddy velocity $\bar{v}_e$ in some direction (peaked at wavenumber $k_0$), the frequency is an angular frequency $\omega_k=\vec{k\cdot \bar{v}}_e$ with $\vec{k}$ the eddy wavenumber. {Then Eq. (\ref{eq-tay-1})} becomes
\begin{equation}
\omega_s=\vec{k\cdot V}+\vec{k\cdot v}_e=kV\Big(\cos\theta_V+\frac{\bar{v}_e}{V}\cos\omega_k t\Big)
\end{equation}
The time average over the fast eddy rotations yields
\begin{equation}
\omega_s=kV\Big(\cos\theta_V+{\textstyle\frac{1}{2}}\frac{\bar{v}_e}{V}\Big)
\end{equation}
implying a velocity dependent correction factor which for large eddy speeds $0<|\cos\theta_V|<1$ dominates.
Th{is} transition from flow to eddy dominated transformation causes a break in the Taylor-transformed spectrum. {Since} $v_e(k)$ depends on wavenumber, the break {will be smoothed.}

{A more subtle observation is that the turbulent frequency $\omega_k$  may depend on wavenumber $\vec{k}$. This ``turbulent dispersion relation''  is  primarily unknown and usually neglected under the assumption that the internal phase velocity of turbulence $\omega_k/k\ll V_0$ is small. It includes quasi-modes,  evanescent oscillations, which in turbulence play the role of ``virtual'' waves\footnote{Their ``virtual'' character  differs from virtual modes in quantum theory where they exist  during time uncertainty $\Delta t<\hbar/\Delta\epsilon_{\vec{p}}$, with $\epsilon_{\vec{p}}=\hbar\omega$, and therefore can never be observed. The turbulent fluctuations of frequency $\omega$ and wavenumber $\vec{k}$ are ``virtual'' in the sense that they are no  solutions of an eigenmode  equation but just cover a set of wavenumbers and frequencies  giving rise to the smooth turbulent wavenumber and frequency spectrum.}, not being waves in real spacetime.  It is unknown whether weak turbulence  theory \citep{yoon2007} can describe it. One way of accounting for them is assuming that internal spatial transport of small scale eddies by large scale eddies at fixed wavenumber $\vec{k}$ causes spectral Doppler broadening \citep{tennekes1975,fung1992,kaneda1993}.}

{Taylor's hypothesis originally referred to stationary, homogeneous, unmagnetised fluid turbulence. The solar wind, on sufficiently large scales, may be treated as a fluid. It is, however, expanding and thus inhomogeneous and nonstationary, and it is magnetised. This raises the question to what degree Taylor's hypothesis does apply to it. As mentioned in the introduction, expansion requires reference to thermodynamics. Thus in application of Taylor's hypothesis one is restricted to local conditions only. In addition the presence of the magnetic field raises the question whether it also holds in application to magnetic fluctuations.}

\section{Magnetic fluctuations in {homogeneous} streaming turbulence}
The magnetic field needs not to be relativistically transformed when dealing with streaming {homogeneous} media. This and its easy observation from spacecraft is its {apparent} advantage over the turbulent electric field $\vec{E}$ which in the observer's frame is given by
\begin{equation}
\vec{E}'=\vec{E}-\vec{V\times B}
\end{equation}
where $\vec{V}$ is the full velocity vector. This well-known relation (which violates the Galilei transformation) causes severe problems in the observation of magnetic turbulence when applying it to the {power} spectra of turbulence. {Applying} Taylor's hypothesis one may consider the turbulent fluctuations $\delta\vec{B}$ of the magnetic field, {expressing them} as Fourier transforms in space and time in the observer's (primed) frame
\begin{equation}
\delta\vec{B}(t',\vec{x}')=\frac{1}{16\pi^4}\int d\omega\,d\vec{k}\:\delta\vec{B}_{\omega\vec{k}} e^{-i\omega t'+i\vec{k\cdot x}'}
\end{equation}
where $\delta\vec{B}_{\omega\vec{k}}$ is a function of frequency $\omega$ and wavenumber $\vec{k}$ with unknown relation between the two.\footnote{One should stress that this representation is completely general. It 
would be wrong to understand the Fourier components as eigenfunctions. The transformation just maps the turbulent fluctuations from real space $\{t,\vec{x}\}\to\{\omega,\vec{k}\}$ into frequency and wavenumber (momentum) space. If there is a relation between $\omega$ and $\vec{k}$ through a turbulent ``dispersion relation'', but to repeat this is by no means a solution of a system of eigenmode equations for the turbulent fluctuations. The system of equations of turbulence is highly nonlinear: it consist of the Fourier-transformed \emph{untruncated} and \emph{unexpanded} dynamical equations of all particle dynamics, flows, and fields.} 
The magnetic field needs not to be transformed. However, in the exponential of its Fourier transform the time and space coordinates are subject to transformation from the turbulence into the observer's frame. 
Transforming these by using Taylor's prescription of using the Galilei transform\footnote{We assume that the total streaming speed $|\vec{V}|\ll c$ though possibly but not necessarily large can be understood as nonrelativistically small. For instance, in the solar wind where Taylor's hypothesis is continuously applied we have \emph{bulk} streaming velocities roughly $200$ km/s $\lesssim V_0\lesssim 2000$ km/s,  occasionally exhibiting very rare peaks which may reach speeds as high as $V_0\sim 4000$ km/s. Hence the ratio $V/c$ is at most in the few percent range allowing to circumvent the complications, reference to relativity would introduce here. However note that in applications to fast expanding stellar winds, like in cataclysmic variables, Wolf-Rayet stars, or even supernova remnants, one would have to refer to relativity.} 
\begin{equation}
\vec{x}=\vec{x'}-\vec{V}t', \quad t'=t, \qquad V\ll c
\end{equation}
the fluctuation becomes
\begin{equation}
\delta\vec{B}(t,\vec{x})=\frac{1}{16\pi^4}\int d\omega\,d\vec{k}\:\delta\vec{B}_{\omega\vec{k}} e^{-i(\omega + \vec{k\cdot V}) t+i\vec{k\cdot x}}
\end{equation}
This suggests that one just has to shift the frequency by the amount $\omega_s=\omega+\vec{k\cdot V}$ which seems simple matter, accounting for the perfect transformation of the wavenumber spectrum of fluctuations into the frequency spectrum in the observers frame. No doubt this holds as long as {the full velocity vector $\vec{V}=\vec{V}_0+\delta\vec{V}$ is known and} the above condition on the angle {for the full velocity} is satisfied, which implies that highly oblique turbulent fluctuations or eddies remain unaffected by Taylor's hypothesis and should be excluded from its application. As long as observations do not strictly distinguish between fluctuations with wavenumbers $\vec{k}\|\vec{V}$ and those with wavenumbers $\vec{k}\perp\vec{V}$, blind application of Taylor's hypothesis {w}ould introduce errors. 

These almost trivial conditions might not seem to be severe but have to be respected (or checked) in any straightforward data analysis. In addition, however, there is another more subtle, interesting, and quite complicated condition which to our knowledge has never been discussed. This we treat separately in the next section.

\section{Velocity dependence}
The neglect of any wavenumber dependence in the turbulent frequencies is a weak {though not too disturbing} approximation in the application of Taylor's hypothesis. A more complicated problem is the dependence on the eddy velocities {$\delta\vec{V}$} which are the building blocks of the turbulence.

The Fourier transform of velocity fluctuations {yields}
\begin{equation}
\delta\vec{V}(t,\vec{x})=\frac{1}{16\pi^4}\int dw\,d\vec{\kappa}\:\delta\vec{V}_{w\vec{\kappa}} e^{-i(w + \vec{\kappa\cdot V}) t+i\vec{\kappa\cdot x}}
\end{equation}
where we used $\vec{\kappa}$ for the wave number of the velocity fluctuations and $w$ for its frequency. 
The full velocity $\vec{V}=\vec{V}_0+\delta\vec{V}$ appears here in the exponential{, which complicates the problem.} 
{Assuming} that in the exponent the turbulent fluctuations are not overwhelmingly important{,} 
we replace $V\to V_0 +\vec{U}_0$, where $\vec{U}_0$ is the mean velocity of the energy-carrying large vortices in the turbulence which affects the evolution of small-scale eddies. This effect of self-interaction in turbulence  remains \citep{tennekes1975}. We then write
\begin{equation}
\delta\vec{V}(t,\vec{x})=\frac{1}{16\pi^4}\int dw\,d\vec{\kappa}\:\delta\vec{V}_{w\vec{\kappa}} e^{-i(w(\vec{\kappa}) + \vec{\kappa\cdot V}_0+\vec{U}_0) t+i\vec{\kappa\cdot x}}
\end{equation}
for the turbulent fluctuations of the velocity. For the moment we will neglect $U_0$ but will revive it below in the context with Doppler broadening.

The  replacement $\vec{V}\to\vec{V_0}$  is, however, not allowed in the Fourier expression of the magnetic fluctuations $\delta\vec{B}$ unless good reasons can be found for its justification. There we have to account for the full speed, the flow plus its turbulent fluctuations{,} in the exponent
\begin{equation}
\delta\vec{B}(t,\vec{x})=\frac{1}{16\pi^4}\int d\omega\,d\vec{k}\:\delta\vec{B}_{\omega\vec{k}} e^{-i\{\omega(\vec{k}) + \vec{k\cdot }[\vec{V}_0+\delta\vec{V}(t,\vec{x})]\}t+i\vec{k\cdot x}}
\end{equation}

The argument that the fluctuations $\delta\vec{V}$ are small for high-speed flows is not a good one, because even for $V_0=0$ in the stationary frame of the turbulence the turbulent magnetic field fluctuations are determined by the fluctuations $\delta\vec{V}$ in the velocity field.  This is another expression for the fact that the electromagnetic field never becomes turbulent by itself. Its fluctuations are always the consequence of turbulence in the electromagnetically active medium. This is taken care of by Maxwell's equations and the dynamical material equations. It is thus  important to recognise that the effect of the velocity turbulence on the magnetic fluctuations has to be taken care of in the Fourier amplitudes even in the frame of turbulence in the absence of flow. 

{The \emph{physical} reason is that the magnetic field fluctuations are \emph{long range}. They correlate with velocity fluctuations over large distances and thus include a substantial part of the mechanical turbulence spectrum.}

Since {$\delta\vec{V}$ is a Fourier integral itself, any}  treatment becomes  involved. One way of dealing with this expression is referring to cumulant expansions \citep[see][for the theory]{kubo1962,fox1976}. This  procedure implies taking the logarithm of the exponential with argument $-i\vec{k}\cdot\delta\vec{V}t$, expanding the exponential in the velocity fluctuation, ensemble averaging over the ensemble of fluctuations assuming  $\langle\delta\vec{V}\rangle=0$, when averaging term by term and rearranging,
\begin{equation}
\big\langle\log\:\exp(-i\vec{k}\cdot\delta\vec{V}t)-1\big\rangle=\sum_{m=1}^\infty\frac{(-1)^m}{(2m)!}{\big\langle(\vec{k}\cdot\delta\vec{V})^{2m}\big\rangle\: t^{2m}}
\end{equation}
This corresponds to a gaussian distribution of the ensemble of velocity fluctuations with zero mean. Any finite mean $\vec{U}_0\neq0$ is added to the mean stream speed. Re-exponentiating  yields the cumulant expansion 
\begin{equation}\label{eq-cumcum}
\delta\vec{B}(t,\vec{x})=\frac{1}{16\pi^4}\int d\omega\,d\vec{k}\:\delta\vec{B}_{\omega\vec{k}} e^{-i[\omega(\vec{k})t -\vec{k\cdot x}]-\frac{1}{2}{\langle(\vec{k}\cdot\delta\vec{V})\rangle^2}t^2+\cdots}
\end{equation}
of which only the  lowest order term in the exponent is retained. This term is negative, quadratic in the average fluctuations and time {$t$} implying a turbulent correlation of the mechanical velocity turbulence and magnetic fluctuations when transported downstream.\footnote{Inclusion of linear wave damping has been considered by \citet{narita2017}.} The square of velocity fluctuations is related to the correlation function of the velocity fluctuations which is the Fourier transform of
\begin{eqnarray}
\delta\vec{V}^2(t',\vec{x}')&=&\frac{1}{(2\pi)^8}\int dw'dw''\,d\vec{\kappa}'d\vec{\kappa}''\delta\vec{V}_{w'\vec{\kappa}'}\delta\vec{V}_{w''\vec{\kappa}''}\nonumber\\
&&\qquad\times\: e^{-i(w'+w'')t'+i(\vec{\kappa}'+\vec{\kappa}'')\cdot\vec{x}'}
\end{eqnarray}
where time $t'$ and space $\vec{x}'$ refer to the time and space dependencies of the turbulent velocity fluctuations. The ensemble averaged velocity spectrum in this case becomes an average over the primed fluctuation scales
\begin{eqnarray}
\langle\delta\vec{V}^2\rangle_{w\vec{\kappa}}(t,\vec{x})&=&\frac{1}{(2\pi)^4\Delta T\Delta V}\nonumber\\
&\times&\!\!\!\!\int dw'\,d\vec{\kappa}'\delta\vec{V}_{w'\vec{\kappa}'}(t,\vec{x})\delta\vec{V}_{w-w',\vec{\kappa}-\vec{\kappa}'}(t,\vec{x})
\end{eqnarray}
the usual result, with $\Delta T, \Delta V$ the corresponding time and volume which are averaged over. This enters the exponent in Eq. (\ref{eq-cumcum}) through its inverse Fourier transform
\begin{eqnarray}
\langle\delta\vec{V}^2\rangle(t',\vec{x}';t,\vec{x})&=&\frac{1}{(2\pi)^8\Delta T\Delta V}\int dw\,d\vec{\kappa}e^{-iwt'+i\vec{\kappa}\cdot\vec{x}'}\nonumber\\[-1ex]
&&\\[-1ex]
&\times&\!\!\!\!\!\!\int dw'\,d\vec{\kappa}'\delta\vec{V}_{w'\vec{\kappa}'}(t,\vec{x})\delta\vec{V}_{w-w',\vec{\kappa}-\vec{\kappa}'}(t,\vec{x})\nonumber
\end{eqnarray}
Note that it may still depend on the observer time and space $t,\vec{x}$. 
For stationary turbulence simplification is achieved by $\delta\vec{V}_{-w,-\kappa}=\delta\vec{V}^*_{w\kappa}$ which allows use of its energy spectrum $\mathcal{E}_{\delta\vec{V}}\propto \langle|\delta\vec{V}|^2\rangle_{w'\kappa'}$, in which case one has
\begin{equation}
\langle\delta\vec{V}^2\rangle(t',\vec{x}';t,\vec{x})\propto\int dw\,d\vec{\kappa}e^{-iwt'+i\vec{\kappa}\cdot\vec{x}'}
\!\!\!\!\!\!\!\!\int dw'\,d\vec{\kappa}'\mathcal{E}_{\delta\vec{V},w'\kappa'}(t,\vec{x})
\end{equation}
These expressions demonstrate the complications introduced when taking into account the effect of mechanical turbulence on the magnetic fluctuations and  ultimately on  magnetic turbulence. 

Complete separation of turbulence from flow implies independence of the  primed and unprimed scales. This holds only approximately because the large-scale eddies in the turbulence, which contain most of the turbulent energy, cause transport of the spectrum of small scale eddies while their own scales approach those of the unprimed flow. They stretch and deform the small-scale eddies. 
Simulations of pure stationary homogeneous velocity turbulence  with $V_0=0$ \citep{yakhot1989,fung1992,kaneda1993} have demonstrated these Doppler broadening effects.  
We will briefly refer to them below in the context of inclusion of a model of turbulence. We will, however, not make use of the above general expression for the cumulative spectral density of the velocity turbulence in the exponential in Eq. (\ref{eq-cumcum}). We rather restrict to Kolmogorov inertial range power law spectra. Indeed, turbulent velocity spectra in the solar wind have been measured \citep[e.g.,][]{podesta2006JGRA,podesta2007}. They exhibit ranges of power laws thus suggesting a continuous spectrum of eddies and vortices in some limited scale range of substantial amplitude and energy content.

\subsection{Simplified case}
In the Appendix we treat the general case. Here let us consider the strongly simplified example, when the turbulent velocity fluctuations are dominated by a narrow wavenumber interval centred around a dominating turbulent eddy $(w_0,\kappa_0)$.\footnote{For simplicity we choose a fixed wavenumber but could as well stay with only a fixed frequency $w_0$ and retain the full $\kappa$ spectrum as only the frequency is subject to the Taylor-Galilei transformation.} Then $\delta\vec{V}_{w\vec{\kappa}}\propto16\pi^4 \delta(w-w_0,\vec{\kappa-\kappa}_0)$. One may think of such a situation as realised in intermittent turbulence for instance in the foreshock \citep{narita2006} and the magnetosheath where single mode eddies seem to be continuously present. They arise from the presence of the collisionless bow shock and flow down the magnetosheath not having had time to decay into a broad band of turbulence. In such a case the whole velocity integral reduces to some complex amplitude 
\begin{eqnarray}
\delta\vec{V}(\vec{x},t)&=& C\:\delta\vec{V}_{w_0\vec{\kappa}_0}e^{-i[(w_0+\vec{\kappa}_0\vec{\cdot V}_0)t-\vec{\kappa}_0\vec{\cdot x}]}\nonumber\\ 
&\sim& C\delta\vec{V}_{w_0\vec{\kappa}_0}\{1-i[(w_0+\vec{\kappa}_0\vec{\cdot V}_0)t-\vec{\kappa}_0\vec{\cdot x}]\}\nonumber
\end{eqnarray}
where for simplicity we expanded the exponential  to just demonstrate the main effect. With $\Delta t$ the time of measurement and $L_\perp$ a transverse dimension of measurement, the factor of proportionality is $C\equiv V_0(L_\perp\Delta t)^2$.  The exponent in the central equality can be lowered by reference to the identity
\begin{displaymath}
e^{-i\psi }= \cos\psi-i\sin\psi
\end{displaymath}
which shows that after integration with respect to $w$ and $\kappa$ the velocity fluctuation will contribute an imaginary and a real part to the exponential in the magnetic fluctuation. The imaginary contribution just shifts the frequency by another amount (which in common applications of the Taylor hypothesis is ignored). The real part introduces some higher order time dependence. As we will show, it implies that the velocity fluctuations, when transformed into the observers frame -- an unavoidable step in any observations --, cause some kind of dissipation in the magnetic fluctuations and thus some kind of deformation of the measured magnetic fluctuation spectrum
\begin{eqnarray}
\delta\vec{B}(t,\vec{x})&=&\frac{1}{16\pi^4}\int d\omega\,d\vec{k}\:\delta\vec{B}_{\omega\vec{k}}\nonumber\\[-1ex]
&&\\[-1ex]
&\times& e^{-i[\omega + \vec{k\cdot }(\vec{V}_0+C\delta\vec{V}_{w_0\vec{\kappa}_0}\{1-i[(w_0+\vec{\kappa}_0\vec{\cdot V}_0)t-\vec{\kappa}_0\vec{\cdot x}]\}) ]t+i\vec{k\cdot x}}\nonumber
\end{eqnarray}
This dissipation is due to the correlations between the different velocity and magnetic scales mentioned above. 
To be as simple as possible here, instead of lowering the exponent, we expand the exponential as shown on the right in the last expression.
This contributes real and complex terms in the exponential argument, second order terms in the time $t$ and a mixed term $t\vec{x}$. The linear real contribution  modifies the oscillating phase shifting it a out of the zero order Taylorian $k_0V_0$ contribution. The imaginary term in $\delta\vec{V}$  causes the noted dissipation in time, an effect of de-correlating  velocity and magnetic field fluctuations and some weak energy loss which might be visible in the magnetic power spectra by small deviations of the slope from Kolmogorov to become slightly steeper, for instance, as is frequently observed.

To show this, assume that $\delta\vec{V}_{w_0\vec{\kappa}_0}^\mathit{real}$ is a real amplitude (in reality it is complex). The exponent in the fluctuation integral becomes  
\begin{eqnarray}
&-&\!\!\!\!i\{[\omega+\vec{k\cdot}(\vec{V}_0+C\:\delta\vec{V}^\mathit{real}_{w_0\vec{\kappa}_0})]t-\vec{k\cdot x}\}\nonumber\\
&&-C\:[(w_0+\vec{\kappa}_0\vec{\cdot V}_0)t^2-\vec{\kappa}_0\vec{\cdot x}t]\vec{k\cdot}\delta\vec{V}^\mathit{real}_{w_0\vec{\kappa}_0}\nonumber
\end{eqnarray}
Define $\tau=t-\vec{\kappa}_0\vec{\cdot x}/w_0'$ with $dt=d\tau$ and $w_0'=w_0+\vec{\kappa}_0\vec{\cdot V}_0$. We also have  $\omega'=\omega + \vec{k\cdot }(\vec{V}_0+C\delta\vec{V}^{\mathit{real}}_{w_0\vec{\kappa}_0})$. With these definitions we quadratically complete the exponents. The turbulent magnetic fluctuations become
\begin{eqnarray}\label{eq-bfluc}
\delta\vec{B}(t,\vec{x})&=&\frac{1}{16\pi^4}\int d\omega\,d\vec{k}\:\delta\vec{B}_{\omega\vec{k}}e^{-C\vec{k\cdot}\delta\vec{V}_{w_0\vec{\kappa}_0}^\mathit{real}[w_0'\tau^2+(\vec{\kappa}_0\vec{\cdot x})^2/4w_0']}\nonumber\\[-1ex]
&&\\[-1ex]
&\times& e^{-i[\omega' \tau+(\omega'\vec{\kappa}_0/w_0'-\vec{k})\vec{\cdot x}]}\nonumber
\end{eqnarray}
which explicitly exhibits the Gaussian damping or decorrelation in the real exponent. (In the Appendix we perform the calculation of the integrals up to the step they can be done analytically.) 

The appearance of decorrelation is solely due to the action of the turbulence. It occurs both in time and space and depends on wavenumber $k$ and the turbulent velocity spectrum. It increases with $k$ and cannot be eliminated when forming the power spectrum. Its presence indicates that the mechanical turbulence has some effect on the wavenumber shape of the spectrum when transformed into the observer's stationary frame.   

One is not interested in the fluctuations but in their Fourier spectrum which is obtained by applying the inverse transform. 
\begin{equation}\label{eq-final}
\delta\vec{B}_{\varpi\vec{K}}=\int dt\,d\vec{x}\:e^{i\varpi t-i\vec{K\cdot x}}\delta\vec{B}(t,\vec{x})
\end{equation}
This yields the replacements for the frequency $\omega'=\varpi$ and for the wavenumber $\vec{K}=(\omega'/w_0')\vec{\kappa}_0-\vec{k}$ which shows that the measured frequency $\varpi\to\vec{k\cdot} (\vec{V}_0+C\delta\vec{V}^\mathit{real}_{w_0\vec{\kappa}_0})$ becomes linearly transformed into the wavenumber space by the sum of the flow and dominant turbulent speeds. This is the slightly varied wanted result. The measured wavenumbers, on the other hand, depend on the wavenumber of the turbulence in a more complicated way. They also include the frequency of the dominant eddy. 
 
The general form of the magnetic fluctuations is obtained when using the exponential identity and separating into real and imaginary parts. This is done in the Appendix. It retains the $w-\vec{\kappa}$ integration of the turbulent velocities and the full trigonometric functions of which the former expression just retains only the first expansion terms. For all practical purposes such an expression becomes incapable because the spectrum of velocity fluctuations is badly known. Thus one stays with the above short term expression, its frequency and wavenumber spectrum. The power spectral density of magnetic fluctuations is obtained from the above expression in the usual way. Measurements usually refer to power spectral densities instead of Fourier spectra. These are obtained in the usual way.

\subsection{Magnetic power spectrum in the simplest case}
{Turbulence theory deals with power spectra. It is not interested in the fluctuations themselves. We therefore proceed to a formulation of the Taylor transformed magnetic power spectrum.}

It is not too difficult to obtain an integral equation for the power spectrum of the magnetic fluctuations in the observers spacecraft frame when just restricting to the above simplest case which includes the velocity effect of turbulent eddies. For this the magnetic field fluctuations in the spacecraft frame must be averaged over time and space. 

Squaring Eq. (\ref{eq-bfluc}) and taking its Fourier amplitude we obtain the turbulent magnetic energy spectrum
\begin{eqnarray}\label{eq-power}
 \big\langle\big|\delta\vec{B}\big|^2\big\rangle_{\omega\vec{k}} &=& \frac{(TV)^{-1}}{4(2\pi)^6}\int \frac{\delta\vec{B}_{\omega_1\vec{k}_1}\delta\vec{B}_{\omega_2\vec{k}_2}\:d\omega_1d\omega_2 d\vec{k}_1d\vec{k}_2}{\sqrt{Cw'_0\delta\vec{V}^\mathit{real}_{w_0\kappa_0}\cdot(\vec{k}_1+\vec{k}_2)}\prod_i[k_{1i}+k_{2i}-k_i-(\omega'_1+\omega'_2)\kappa_{0i}/w'_0]}\nonumber\\
&\times&\exp\Big\{-\frac{(\omega'_1+\omega'_2-\omega)^2+\sum_i\kappa_{0i}^{-2}[(k_{1i}+k_{2i}-k_i)-(\omega'_1+\omega'_2)\kappa_{0i}/w'_0]^2}{4[Cw'_0\delta\vec{V}^\mathit{real}_{w_0\kappa_0}\cdot(\vec{k}_1+\vec{k}_2)]}\Big\}
\end{eqnarray}
\noindent
where $\omega'_{1,2}=\omega_{1,2} + \sum_ik_{1,2i}V'_{0i}$, with $\vec{V}'_0=\vec{V}_0+C\delta\vec{V}^{\mathit{real}}_{w_0\vec{\kappa}_0}$. Here $T$ and $V$ are the respective time interval and volume over which the time and space averages are taken. These are determined by the experimental condition{s} and instrumental resolutions.

The triple product in the denominator implies the presence of three singularities in the integral which can be exploited for its simplification. The assumption of  absence of singularities in the spectra themselves implies that they are entire functions. For this to hold the turbulence is free of any eigenmodes. Otherwise one {deals with intermittency and} needs to include the{ir} residua. This {must} only be done if they have been identified in the observations. Unfortunately, their non-identification does not imply their absence as they may be hidden in the overall spectrum of their frequency range {but are not} resolved.

The singularities are in the three components of  wavenumbers or frequencies. To see this we rewrite the general term in the product as
\begin{equation}
k_{1i}+k_{2i}-k_i-(\omega'_1+\omega'_2)\kappa_{0i}/w'_0 = k'_{1i}+k'_{2i}-k_i-(\omega_{1}+\omega_{2})\kappa_{0i}/w'_0
\end{equation}
with the definition $\vec{k}'_{1,2}\equiv \big(\mathbf{I}-\vec{\kappa}_{0}\vec{V}'_{0}/w'_0\big)\cdot\vec{k}_{1,2}$. This yields the wavenumber volume element 
\begin{equation}
d\vec{k}'_1d\vec{k}'_2=\Big(\mathbf{I}-\frac{\vec{\kappa}_0\vec{V}'_0}{w'_0}\Big)^2:d\vec{k}_1d\vec{k}_2
\end{equation}
which must be inverted in order to be able to replace the unprimed volume element. It produces the factor $\big(\mathbf{I}-\vec{\kappa}_0\vec{V}'_0/w'_0\big)^{-2}$ under the integral sign, which is the scalar square product of the inverse tensor. Integrating with respect to $\vec{k}'_2$ over the upper complex plane and accounting only for the resonant part, which assumes that the magnetic fluctuation spectra $\delta\vec{B}_{\omega_2\vec{k}_2}$ have no singularities neither in this plane nor on the real axis, a rather strong restriction by itself (in principle neglecting resonant wave-wave interactions), produces a factor $(i\pi)$ in front of the integral while replacing the resonant product in the denominator by 
\begin{equation}
\delta\Big(k'_{2i}+k'_{1i}-k_i-\Omega_{1i}-\Omega_{2i}\Big)
\end{equation}
in the numerator, while leaving in the denominator a product of the two remaining components $j\neq i$. It is this factor which in the integration with respect to the $i$th component of the $\vec{k}'_2$ introduces a mixing of indices when eliminating this component of the second wavenumber. In this way the integral becomes the sum of three integrals. 

Let us define $A^{-1}=1-\vec{\kappa}_0\cdot\vec{V}'_0/w'_0$. Then we have at resonance
\begin{eqnarray}
&&k_{2i}=-k_{1i}+A[k_i+\kappa_{0i}(\omega_1+\omega_2)/w'_0]\nonumber\\
&&\omega'_1+\omega'_2=(\omega_1+\omega_2)(1+A\vec{V}'_0\cdot\vec{\kappa}_0/w'_0)+A\vec{V}'_{0}\cdot\vec{k}\\
&&\delta\vec{V}^\mathit{real}_{w_0\kappa_0}\cdot(\vec{k}_1+\vec{k}_2)=A\delta\vec{V}^\mathit{real}_{w_0\kappa}\cdot[\vec{k}+\vec{\kappa}_{0}(\omega_1+\omega_2)/w'_0]\nonumber
\end{eqnarray}
These expressions have to be used in Eq. (\ref{eq-power}). They enter the denominator under the root and the fraction in the gaussian exponential. Unfortunately their appearance in these places in mixed form containing both frequencies $\omega_{1,2}$ inhibits any further analytical treatment even in the resonant simplified case.
Moreover, the indices of the magnetic fluctuation spectra become affected, changing to
\begin{equation}
\delta\vec{B}_{\omega_2\vec{k}_2}\to\delta\vec{B}_{\omega_2, [\vec{k}-\vec{k}'_1+(\omega_1+\omega_2)\vec{\kappa}_0/w'_0]}
\end{equation}

Similar difficulties arise if the resonances are attributed to one of the frequencies $\omega_{1,2}$. In this case, defining frequencies $\Omega_i=w'_0k_i/\kappa_{0i}, \vec{k}=\Omega\vec{\kappa}_0/w'_0$, the power spectrum of the magnetic energy density is subject to the equation
\begin{eqnarray}
\big\langle\big|\delta\vec{B}\big|^2\big\rangle_{\omega\vec{k}} &=& \frac{i}{(4\pi)^4TV}\sum_i\int \frac{\delta\vec{B}_{\omega_1\vec{k}_1}\delta\vec{B}_{(\omega'_1+\Omega_i-\Omega_{1i}-\Omega_{2i})\vec{k}_2}\:d\omega_1 d\vec{k}_1d\vec{k}_2}{\sqrt{Cw'_0\delta\vec{V}^\mathit{real}_{w_0\kappa_0}\cdot(\vec{k}_1+\vec{k}_2)}\prod_{j\neq i}[k_{1j}+k_{2j}-k_j-\omega'_1\kappa_{0j}/w'_0]} \nonumber\\
&\times&\exp\Big\{-\frac{(\omega-\Omega_i+\Omega_{1i}+\Omega_{2i})^2+\sum_{j\neq i}\kappa_{0j}^{-2}[k_{1j}+k_{2j}-k_j-\omega'_1\kappa_{0j}/w'_0]^2}{4[Cw'_0\delta\vec{V}^\mathit{real}_{w_0\kappa_0}\cdot(\vec{k}_1+\vec{k}_2)]}\Big\}
\end{eqnarray}
which is the sum of three integrals. Here the volume element becomes 
\begin{equation}
d\vec{k}_1d\vec{k}_2=(\kappa_{01}\kappa_{02}\kappa_{03})^2d\Omega_{11}d\Omega_{12}d\Omega_{13}d\Omega_{21}d\Omega_{22}d\Omega_{23}/w'^6_0 
\end{equation}
In addition $\omega'_1=\omega_1+\vec{k}_1\cdot\vec{V}'_0$ has to be replaced, and the wavenumbers must be expressed through the $\Omega$s. 

This sum of integral equations looks simpler but is at the best ready for an iterative solution, because the complicated index prevents from combining the fluctuations in the integrand into a power spectrum. This is a simple consequence of the Wiener-Khinchin theorem which has been applied here several times.  The discouraging observation is that, though it seems that the $\omega_1$ integration could be separated, in all and even this simplest case one cannot simply refer to the power spectral energy density of the observed magnetic fluctuations in order to reconstruct the wavenumber spectrum. One needs to solve a set of complicated integral equations which contains the folding product of the magnetic fluctuations in frequency and wavenumber space. Even separating the $\omega_1$ integration requires a shift in one of the fluctuation spectra imposed by the second index in order to perform the folding. The presence of a turbulent fluctuation spectrum in velocity thus implies correlations in the magnetic fluctuations.   
 
Except for an iteration procedure using measured magnetic fluctuation spectra, the only way, even in this most simplistic case under the most simplifying assumptions is to assume a reasonable model for the velocity fluctuations. Imposing such models for the magnetic fluctuation spectrum one expects that the integral can be iteratively solved and the approximate magnetic power spectral density in wavenumber space can be constructed from the measurement of the frequency spectrum. 

What the above expressions show is not only that different spectral domains in frequency space are correlated. It also shows that the spectrum becomes folded by the weight of a fairly complicated gaussian distribution in frequency and wavenumbers, which appears as the kernel of the above integral equations (\ref{eq-power}). Considering the resonant denominator simplifies it slightly but does not release from the necessity of solving them.

The general case of turbulence in the velocity and its stepwise reduction to the Taylor hypothesis is treated in the Appendix. It shows that the Taylor hypothesis concerning the transformation of the magnetic turbulence spectrum is nothing else but the equivalent to the complete neglect of everything in mechanical turbulence except the pure streaming velocity. Whether this in MHD turbulence can be justified, is questionable, because the magnetic and mechanical fluctuations are intimately related and should not be considered separately. Below we return to this point. 

{There is, however, a further simplification for the particular case of purely alfv\'enic turbulence. In this case the linear magnetic fluctuations $\delta\vec{B}\propto\delta\vec{V}$ and the magnetic spectra in the integrand can be expressed through the corresponding spectral fluctuations of the velocity. This results in an expression of the magnetic power spectral density solely through the spectral densities of the velocity fluctuation. This substantially simplifies the transformation. Unfortunately it does not resolve the correlation problem which remains as that of the velocity fluctuations. Nevertheless, the problem reduces to the knowledge of the latter and to an appropriate solution of the singular correlation integral. Ultimately this can  numerically be obtained.}  

\section{Implications to spacecraft data analysis}

\subsection*{Mapping problem}

\begin{figure}[htb]
  \center
  \includegraphics[width=0.85\textwidth]{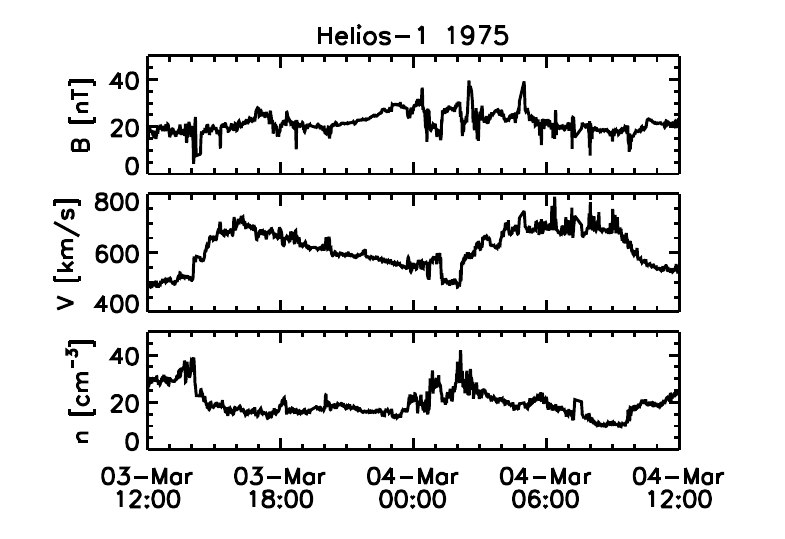}
  \caption{
    {
    Time series plots of the magnetic field magnitude, 
    flow velocity magnitude (for protons),
    and proton number density from the Helios-1 spacecraft measurement
    in the inner heliosphere (at a radial distance of about 0.4 AU
    from the Sun) on March 3--4, 1975.
   }
  }
  \label{fig:obs_timeseries}
\end{figure}

{
  Taylor's hypothesis breaks down when
  fluctuations are no longer negligible in the flow velocity.
  One of the possible ways to overcome the effect
  of the flow velocity fluctuation is to map
  the time series data onto the spatial domain
  in the stream-wise direction (along the flow)
  by correcting for the instantaneous or 
  individual realizations of the flow velocity fluctuation.
  Here we sketch a more appropriate mapping method 
  (than the use of Taylor's hypothesis) along with
  a data analysis for the Helios-1 plasma and 
  magnetic field measurements.
  Figure\,\ref{fig:obs_timeseries} displays
  the magnetic field magnitude, the flow velocity magnitude
  (for protons), and the number density (for protons)
  from Helios-1 spacecraft from March 3, 1975, 1200 UT
  to March 4, 1975, 1200 UT in the time series style.
  The Helios-1 spacecraft was located at a distance
  of about 0.4 AU (Astronomical Unit) from the Sun.
}

\begin{figure}[htb]
  \center
  \includegraphics[width=0.85\textwidth]{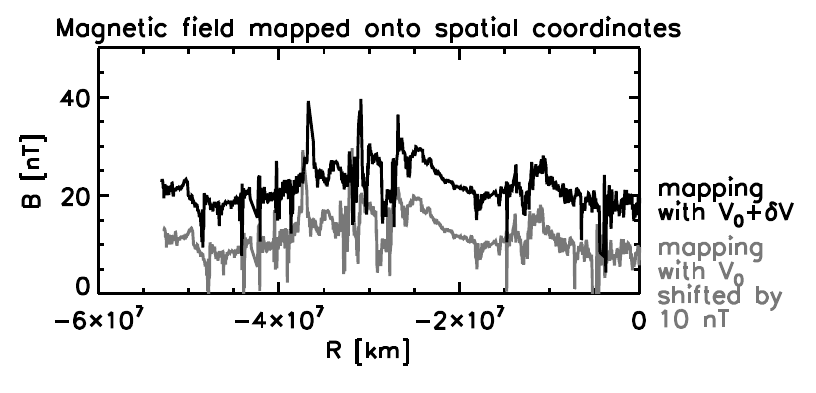}
  \caption{
    {
    Magnetic field data (only the magnitude is plotted)
    mapped onto the streamwise spatial coordinates 
    using the total flow velocity in black (including the
    mean constant flow velocity field $V_0$ and the fluctuating 
    field $\delta V$ and using the mean constant flow velocity
    in gray (Taylor's hypothesis).
    Only the radial direction away from the 
    Sun is considered here. 
   }
  }
  \label{fig:obs_mapped_data}
\end{figure}

{
  The instantaneous, fluctuation-mapping of the spacecraft data 
  from the time domain onto the streamwise spatial domain is obtained by
  the following relation:
\begin{equation}
  R^\mathrm{(map)}(t) = \int_{t_0}^t V(t^\prime) \; dt^\prime 
  \label{eq:mapping}.
\end{equation}
  Here, for simplicity, we consider the radial direction from the Sun
  and use the radial flow component $V_r$ for the mapping.
  The five-point Newton-Cotes algorithm is implemented
  in the numerical integration in Eq.\,\ref{eq:mapping}.
  Positive values in the mapped distance $R$ are
  associated with the streamwise (or anti-sunward) direction 
  (in the observer's frame), and negative values
  are associated with the sunward direction.
  For comparison, the conventional 
  radial mapping under Taylor's hypothesis reads
\begin{equation}
  R^\mathrm{(TH)}(t) = V_0 (t - t_0)
  \label{eq:mapping}.
\end{equation}
  Magnetic field data mapped in the two ways
  (fluctuation-corrected way and Taylor's hypothesis)
  are displayed in Fig.\,\ref{fig:obs_mapped_data}
  as a function of the radial distance from the
  spacecraft position toward the Sun as
  $B(R^\mathrm{(map)})$ in black and $B(R^\mathrm{(TH)})$ in gray.
  The mean flow speed is about 612 km/s in the radial direction
  (away from the Sun).
  The fluctuation-mapped data (in black) apparently have
  a very close waveform to the Taylor-mapped data (in gray), 
  but the two mapped data differ in the spatial positions.
  For example, the peak at $R=-3.7\times 10^{7}$ km in the 
  fluctuation-mapped data is displaced to $R=-3.8\times 10^{7}$ km
  in the Taylor-mapped data, or the waveform is displaced 
  in the opposite order such as the field 
  decrease around $R=-4.8 \times 10^{7}$ km.
  The fluctuation-based mapping in the turbulence observation 
  from the time domain onto the spatial domain
  may thus be regarded as a shuffling of the data
  without changing the statistics or the 
  probability distribution function of the fluctuations.
}

\begin{figure}[htb]
  \center
  \includegraphics[width=0.85\textwidth]{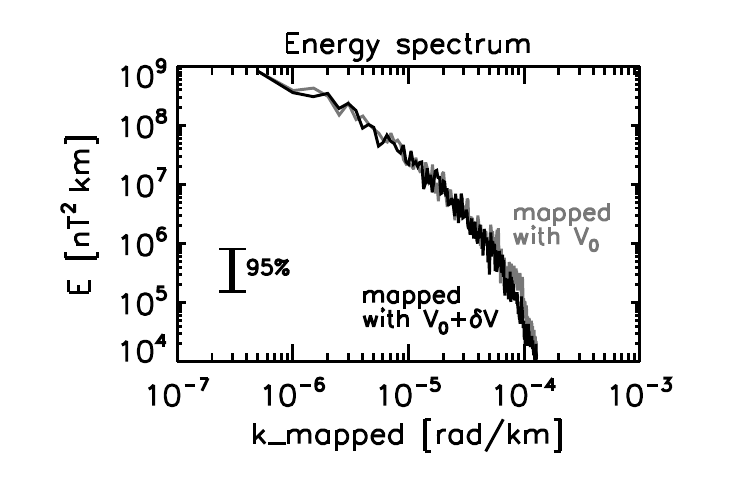}
  \caption{
    {
    Energy spectrum (trace of the spectral density matrix)
    of the mapped magnetic field data (onto the spatial coordinates)
    in the streamwise wavenumber domain. The spectrum
    using the total flow velocity is represented in black,
    and that using the mean constant flow velocity
    is in gray.
   }
  }
  \label{fig:obs_spectra}
\end{figure}

{
  Energy spectrum for the magnetic field 
  is compared between the fluctuation-mapped data
  and the Taylor-mapped data (Fig.\,\ref{fig:obs_spectra}).
  The fluctuation-mapped data are irregularly
  displaced in the spatial domain, and are
  re-sampled into a regular sampling data set by interpolation. 
  The spectrum is evaluated by the Welch-FFT 
  (fast Fourier transformation) algorithm
  with a window size of 512 data points,
  a sliding of 128 data points, and
  12 degrees of freedom (which is the number 
  of sub-intervals for the statistical averaging)
  Figure\,\ref{fig:obs_spectra} displays
  the total fluctuation energy (in the spectral domain)
  over the three components of the magnetic field
  (which is the trace of the spectral
  density matrix) as a function of the streamwise
  wavenumbers for the fluctuation-mapped data
  (in black) and the Taylor-mapped data (in gray).
  The fluctuation energy has nearly the same
  spectral shape in the lower wavenumber range
  up to about $5 \times 10^{-5}$ rad km$^{-1}$.
  The spectrum becomes gradually and increasingly 
  steeper in the fluctuation-mapped data
  at about $10^{-4}$ rad km$^{-1}$, while
  the spectrum exhibits a break at about $10^{-4}$ rad km$^{-1}$
  and becomes suddenly steeper in the Taylor-mapped data. 
  Therefore, the use of the Taylor's hypothesis
  may introduce a spectral deformation
  when the fluctuation in the flow velocity is not negligible.
}

\subsection*{Fluid-like or MHD-like breakdown of Taylor's hypothesis}

{
  Time dependence of magnetic field fluctuation in the observer's frame 
  consists of the three distinct factors: the MHD-intrinsic fluctuation 
  (e.g., Alfv\'en wave) with the frequency $\omega = k_\| V_A$,
  the advection by the mean flow velocity or the Doppler shift $k V_0$,
  and the random sweeping by the fluctuating flow velocity $\omega$  as follows.
  \begin{equation}
    \delta B(t) \propto 
      \exp\left[
        -i \omega t - i k V_0 t
        - i k \delta V t
      \right]
    \label{eq:three_factors}.
  \end{equation}
  Taylor's hypothesis breaks down 
  either under the finite intrinsic frequencies (of the Alfv\'en waves)
  as $k_\| V_A $ or under the finite fluctuation amplitude in the flow velocity
  as $k \delta V$. We associate the fluctuating flow velocity 
  with the perpendicular wavenumbers (representing eddies
  around the mean magnetic field), and can derive an estimate
  of the sense of the breakdown of Taylor's hypothesis
  (fluid-like or MHD-like) by taking a ratio of the two frequency quantities,
  \begin{eqnarray}
    r &=& 
      \frac{k_\perp \delta V}{k_\| V_a}
      \label{eq:ratio01}\\
      &=& 
      \frac{\delta V}{V_A \tan\theta}
      \label{eq:ratio02},
  \end{eqnarray}
  where the angle $\theta$ is defined as $\tan \theta = \frac{k_\|}{k_\perp}$
  and is approximated to the angle between the mean flow
  and the mean magnetic field in the application to the observation,
  $\theta \simeq \theta_{V_0 B}$. Thus the solar wind observation 
  in a more radial mean magnetic field from the Sun 
  (a small value of $\tan \theta_{V_0 B}$)
  is influenced by the fluid-like breakdown of Taylor's hypothesis
  (inaccurate mapping onto the spatial coordinates),
  and that in a more perpendicular mean magnetic field 
  to the direction from the Sun (a large value of $\tan \theta_{V_0 B}$)
  is influenced by the MHD-like breakdown of Taylor's hypothesis
  (intrinsic Alfv\'en waves and counter-propagating Alfv\'en waves).
}


\section{{Referring to Kolmogorov's turbulence spectrum}}

So far (including the Appendix) we did not refer to any model of the turbulence. The entire approach given was just from the point of view of the observer {who  measures fluctuations and does not primarily ask for a model}. It illuminates the purely Galilean effect of the transport of turbulence across the observers frame and the prospects of accounting for the presence of turbulence in the reduction of the magnetic power spectrum. 
{G}iven any turbulence, no matter how it is generated and evolves, the spectrum of 
magnetic fluctuations swept across the observer has been considered as to its modification by the turbulent fluctuations in the velocity {field}. From that point of view our endeavour (as completely given in the Appendix) is rather general. {T}here have been attempts in the literature to approach the Taylor problem from the side of turbulence theory. In these cases theory {\emph{provides}} a theoretical spectrum of turbulence which then is 
{set} to flow. As far as such theoretical spectra of the turbulent flow are concerned, they enter into our turbulent velocity terms $\delta\vec{V}_{w\vec{\kappa}}$ or their squared time averages $\langle|\delta\vec{V}|^2\rangle$. 

Among the turbulence models the most prominent are Kolmogorov's \citep{kolmogorov1941a,kolmogorov1941b,kolmogorov1962} and, from the point of view of  turbulence in a magnetised medium like MHD,  Iroshnikov and Kraichnan's \citep{iroshnikov1964,kraichnan1965,kraichnan1967PhFl} models both discussed widely in the literature \citep[cf., e.g.,][]{biskamp2003}.

{F}ocus has {also} been given to the energy spectrum of the velocity fluctuations \citep{fung1992,kaneda1993} which in Eq. (\ref{eq-cumcum}) enters through the cumulant expansion term in the exponent of the magnetic fluctuations \emph{before} the magnetic energy spectrum is calculated. The assumption there is that the original mechanical turbulent energy spectrum to start with 
\begin{equation}
\mathcal{E}(\kappa)=\int \mathcal{E}(\kappa, w_\kappa) dw_\kappa 
\end{equation}
is either Kolmogorov $\mathcal{E}_K\propto\epsilon^\frac{2}{3} \kappa^{-\frac{5}{3}}$ or Iroshnikov-Kraichnan $\mathcal{E}_\mathit{IK}\propto(\epsilon V_A)^\frac{1}{2} \kappa^{-\frac{3}{2}}$, with $V_A$ Alfv\'en speed. It becomes deformed by advection
(by the large-scale energy-carrying eddies). Such theoretical and numerical attempts restrict to the inertial range of the velocity turbulence. If the energy in the velocity fluctuations  at  frequency $w_\kappa$  has gaussian spread in wavenumber, the advected  Kolmogorov energy spectrum of the small-scale velocity fluctuations  at mean large-eddy velocity $U_0$ assumes the form \citep{fung1992}
\begin{equation}\label{eq-kolm}
\mathcal{E}^K_{\delta\vec{V}}(\kappa,w_\kappa)\propto\frac{\mathcal{E}_K}{2\kappa U_0}\sum_{\pm}\exp\Big(-\frac{1}{2}\frac{w^2_{\kappa\pm}}{\kappa^2 U_0^2}\Big), \qquad w_{\kappa\pm}=w_\kappa \pm \lambda\epsilon^\frac{1}{3}\kappa^\frac{2}{3}
\end{equation}
Here $\lambda\sim O(1)$ is some constant. The advected Kolmogorov velocity spectrum of the mechanical turbulence thus should steepen for the assumed high large-scale speeds $U_0$
\begin{equation}
\mathcal{E}^K_{\delta\vec{V}}(\kappa,w_\kappa)\propto \kappa^{-\frac{8}{3}}
\end{equation}
At large $\kappa$ the exponential factor in this range tends to unity. At small $\kappa$ it suppresses the spectrum exponentially. Advection thus \emph{{reduces}} the effect of small-scale eddy turbulence on the large eddies. It is the short scales in the inertial range which cause the main deformation of the advected spectrum. 
Taylor's assumption then implies that the frequency of the mechanical turbulence simply becomes $w_\pm\sim \kappa U_0$,  i.e. determined by the speed of the largest eddies. The  $\kappa$-dependent second term in $w_\pm$ is neglected, and the exponential reduces to a number. At high Reynolds numbers and large $U_0$ the internal turbulent dispersion plays no role in mechanical Kolmogorov turbulence. 
Even if  it exists, it is not taken into account anywhere. The advected  Kolmogorov velocity spectrum can  be integrated \citep{tennekes1975} with respect to wavenumber $\kappa$ to become 
\begin{equation}
\mathcal{E}_w^K\sim (\epsilon U_0)^\frac{2}{3}w^{-\frac{5}{3}} 
\end{equation}
which, as expected, is a simple mapping of the velocity spectrum into frequency space or vice versa from frequency into wavenumber space. {This applies to the original frame of turbulence. It does not yet apply to the Taylor-Galilei transformation from the stationary turbulence frame via the large-scale streaming into the observer's frame. The} functional dependence of the inertial range at large {advection} speeds $U_0$ has been reproduced by numerical simulations \citep{fung1992,kaneda1993}. In fact, in order to be somewhat more precise, it should be noted that the complete reduced frequency spectrum obtained \citep{fung1992,kaneda1993} consists of two terms
\begin{equation}\label{eq-wpm}
\mathcal{E}_w= a\,w^{-2}+b\,w^{-\frac{5}{3}}, \qquad a,b\in \{C\} 
\end{equation}
which generalises the former theory \citep{tennekes1975} to inclusion of a large range of Reynolds numbers in the velocity turbulence. Here $a,b$ are $\epsilon$-dependent constants ($c$-numbers). It has, however, been shown \citep{fung1992,kaneda1993} that the first of these terms is always small compared with the second as long as one stays in the inertial range.

Use of these expressions in Eq. (\ref{eq-cumcum}) and applying the $w$-integration yields a delta function, and the average over the velocity turbulence reduces to an additive term $\propto Rk^2w^{-\frac{5}{3}}t^2$ in the exponential with some factor $R$, which depends on $U_0$ and $\epsilon$. (If it depends on time, this dependence is given by $R(U_0t=x-x_0)$ as a mere translation.) The term resulting in the argument of the exponential however retains the irreducible second order time dependence in the exponential as long as the velocity spectrum is not completely neglected. {\emph{Reference to the Kolmogorov spectrum and its mapping thus does not eliminate its effect on the spectrum of magnetic turbulence.}}
  
Whether neglect of the internal frequencies and their dispersion, the turbulent dispersion relation, is justified or not remains an unresolved problem. The large speed simulations seem to justify it at least for the limited inertial range, however on the expense of rather short inertial ranges of less than an order of magnitude in frequency or so obtained in the simulations in the high Reynolds number limit. This suggests that in the inertial range possibly no susceptible dispersion can experimentally be detected and thus in the transformation of the velocity spectrum play{s} no role. The small scale velocity eddies seem  ``frozen'' \citep{fung1992,kaneda1993} in the Kolmogorov transport of energy down the inertial range. 

{This does not resolve the complications in determining the magnetic fluctuation spectrum, however. It just tells that the inertial range velocity spectrum can be transformed but in the magnetic fluctuations must be retained at least in the form of the cumulant expansion term.}

We can, however, understand Eq. (\ref{eq-wpm}) as an inertial range turbulent dispersion relation. The internal turbulent dispersion in the inertial range is given by 
\begin{equation}
w_\kappa = \pm \lambda\epsilon^\frac{1}{3}\kappa^\frac{2}{3}
\end{equation}
which in fact follows directly from inspection of the Kolmogorov spectrum in one dimension as the inverse time-scale in the inertial range. The same argument applied to an Iroshnikov-Kraichnan spectrum $\mathcal{E}_\mathit{IK}$ with $\delta z_\pm\sim  (\epsilon V_A\ell)^\frac{1}{4}$ leads to an isotropic weakly increasing with wavenumber turbulent dispersion relation 
\begin{equation}
w_\kappa\sim (\epsilon V_A)^\frac{1}{4} \kappa^\frac{3}{4}
\end{equation}
in the inertial range interval $\kappa_{in}\ll\kappa\ll\kappa_d$ between the eddy wavenumbers of energy injection $\kappa_\mathit{in}$ and energy dissipation $\kappa_d$. (Note that the latter is not necessarily Kolmogorov's genuine dissipation scale but might simply be the transition from the magnetised electron scale to the domain of scales where the electrons become nonmagnetic and processes take over in which the magnetic field and its fluctuations are not anymore included but electrostatic processes start dominating.) Both these dispersion relations which follow from straight dimensional analyses are weakly nonlinear only. They show that the turbulent frequencies increase with decreasing spatial scale of the velocity eddies, a behaviour reminding of sound waves. Smaller eddies oscillate at larger frequency, an effect of their decreased inertia. These expressions do not account for any anisotropy and higher dimensionality in the velocity turbulence, however, which is justified at large eddy scales in the MHD range but becomes questionable in the Hall and electron-MHD range where the ions demagnetise thus becoming about independent of the magnetic field. Their dependence on the magnetic field is only via their charge neutralising and electric current coupling to the magnetised electrons. Instead of Alfv\'en waves the relevant waves are kinetic Alfv\'en waves \citep{baumjohann1996} with dispersion 
\begin{equation}
w^2(\kappa_\|,\kappa_\perp)=\kappa_\|^2V_A^2\big(1+{\textstyle\frac{1}{2}}\beta\kappa_\perp^2\rho_i^2\big)\big(1+\kappa_\perp^2\lambda_e^2\big)^{-1}
\end{equation}
with $\rho_i$ the ion gyroradius, $\lambda_{e,i}=c/\omega_{e,i}$ electron and ion inertial lengths, and $\omega_{e,i}$ electron and ion plasma frequency, respectively. $\lambda_i>\kappa_\perp^{-1}>\lambda_e$ holds in this range.  Any turbulence in the wavenumber range $\kappa>\rho_i^{-1}$ becomes anisotropic. Its spectrum then splits into parallel and perpendicular components  \citep{goldreich1995}
\begin{equation}
\mathcal{E}_{\kappa_\|} \propto \epsilon^\frac{3}{2}\big(V_A\kappa_\|\big)^{-\frac{5}{2}}, \qquad \mathcal{E}_{\kappa_\perp} \propto \epsilon^\frac{2}{3}\kappa_\perp^{-\frac{5}{3}}
\end{equation}
which in the perpendicular direction is Kolmogorov while in the parallel direction is steeper. This is because the two scales are different \citep{biskamp2003}: $\kappa_\perp/\kappa_\|\sim \big(\sqrt{\beta/2}\rho_i\kappa_\perp\big)^\frac{1}{3}$ or  $\kappa_\perp/\kappa_\|\sim \big(\lambda_i\kappa_\perp\big)^\frac{1}{3}$, which basically is the restriction on $\kappa_\perp$. The deformation of the spectrum then depends on the direction of the main flow parallel or perpendicular to the mean magnetic field. For parallel flow only the parallel spectrum of eddies becomes deformed, while for perpendicular convection of the turbulence it is the perpendicular spectrum which deforms. The former is non-magnetised, while the latter is subject to the Kolmogorov kind of deformation referred to above in Eq. (\ref{eq-kolm}). We than have for $\delta z_\perp\sim\ell_\perp^\frac{1}{3}/V_A$ and $\delta z_\|\sim \ell_\perp/\ell_\|$ which gives the two dispersion relations
\begin{eqnarray}
w_{\kappa_\perp} &\sim& \frac{\epsilon^\frac{1}{3}}{V_A}\kappa_\perp^\frac{2}{3},\qquad (\kappa_\|~~~ \mathrm{fixed})\nonumber\\
&&\\[-3ex]
w_{\kappa_\|} &\sim& \Big(\frac{\epsilon}{V_A^3}\Big)^\frac{1}{2}\kappa_\|^\frac{1}{2},\qquad (\kappa_\perp ~~\mathrm{fixed})\nonumber
\end{eqnarray}
where the parallel relation is just a consequence of the perpendicular dispersion relation.
Application of the Taylor hypothesis to the mechanical turbulence power spectra in these cases is justified as this deformation can for large $U_0$ be mapped to frequency space. This holds for the Kolmogorov and Iroshnikov-Kraichnan spectra as well as their anisotropic extension into the range of Hall scales respectively ion-inertial scales. It implies neglect of the internal frequency dependence of the power spectral densities. However this is of little help in the transformation of the magnetic spectra as these still depend on the presence of the velocity fluctuations respectively mechanical turbulence. Reference to those turbulent power spectral densities does not eliminate this dependence.  

\section{Conclusions}

The above brief investigation shows a number of interesting points which usually are neglected by experimentalists and theorists as well or considered to be simple and not worth any discussion. So straight recalculations of power spectral densities recorded at spacecraft from frequency into wavenumber space adopting Taylor's hypothesis are common wisdom. 

The above  analysis demonstrates that such a simple replacement is possible in the Fourier spectra of the magnetic fluctuations, though under some listed rather severe restrictions. 

We do not want to be as rude as \citet{saint2000} concerning the application of the Taylor hypothesis to observations of power spectral densities of magnetic turbulence in a streamin plasma like the solar wind in order to obtain an impression of some part of the wavenumber spectrum of the turbulence, where it can be applied, in particular in the more frequently realised case when no direct measurements of the velocity spectrum or its wavenumber distribution are available. Monitoring  magnetic fluctuations is easiest and has the advantage of obeying relativistic invariance. However, when the turbulent velocity fluctuations have to be taken into account because they become comparably large amplitude and thus cannot be ignored, as should be the case in slow solar wind flows for instance, then the problem of the required transformation into the spacecraft frame becomes more complicated. 

The above considerations (including those in the Appendix) demonstrate quite clearly that the restriction to observation of the mechanical turbulence, viz. the velocity (and also the density) fluctuations and their power spectral densities permits application of the Taylor hypothesis. This permission is granted for high Reynolds numbers and fast flows and applies to the inertial range. We have used this in discussing the Kolmogorov spectrum for homogeneous and isotropic turbulence of the flow. We have not yet considered the effects of anisotropies in the flow which in MHD must naturally occur because of the differences in the flow parallel and perpendicular to the magnetic field, i.e. the free flow along a mean external field and its convection perpendicular to this mean field. Even in this case one must distinguish between strong and weak field conditions as the relation between flow and field differ in those cases. Aside of these cases, in the homogeneous and isotropic turbulent flow the inertial range turbulence suggests that the large energy-carrying eddies freeze the small-scale eddies, and there is a range in which the spectrum can indeed be simply mapped from frequency space into wavenumber space and vice versa. Thus here the Taylor hypothesis applies under conditions of large Reynolds numbers and some gap between large and small eddy turbulence.

However, when from this step going up the ladder to the magnetic power spectral densities becomes a rather more difficult endeavour. The full turbulent velocity spectrum appears in the argument of the exponential in magnetic fluctuations. The transformation of the turbulent flow into the spacecraft frame changes the frequency mapping into the wavenumber spectrum. There is no linearity between the two as this is distorted even in the simplest cases. Only under severe restrictions (as identified in the Appendix) this straight mapping from mechanical into magnetic turbulence can be successfully done. In the general case even estimates of the different ranges of  plasma parameters like the wavenumbers corresponding to gyroscales and inertial scales must  be cautioned. The Taylor hypothesis applied to magnetic turbulence should be restricted to fast flows only, substantially faster than the fastest expected rotational velocities of the turbulent eddies in the mechanical flow. Such conditions are given  in high speed solar wind or stellar outflows.  Except for the wavenumber and frequency dependent ``damping effect'' of the spectrum found here, this has been, in principle, all well known already. 

In MHD the magnetic and mechanical turbulence are intimately connected, which suggests use of the well-known alfv\'enic Elsasser variables $z_\pm$ \citep{elsasser1950}. At high Reynolds numbers and weak kinetic-magnetic correlations this leads to Iroshnikov-Kraichnan spectra \citep[cf., e.g.,][for a review]{biskamp2003} for the subset of a non-streaming broad spectrum of turbulent eddies to which we referred above. At the shorter, non-alfv\'enic scales below MHD scales in magnetic turbulence in the Hall- or electron-MHD range of a non-magnetised ion  and magnetised electron fluids, reference to kinetic Alfv\'en rather than Alfv\'en waves is more appropriate. They naturally introduce an anisotropy through the appearance of the inertial length as a natural scale in this range and internal transport mainly along the mean magnetic field.

{
  Another important consideration is whether or not one can even, 
  in principle, deduce the wavenumber from the time series. 
  For example, again from the Helios data, when the spectral index is $f^{-1}$ 
  (in the frequency domain) one has to check if the implied wave length 
  is less than the distance of the spacecraft to the Sun. 
  If it isn't, Taylor's hypothesis won't help. 
  So, it is interesting to address the question, ``can we demonstrate 
  instances where their theoretical analysis makes  a quantitative difference?''
  Naively speaking, the Task of the wavenumber determination from the 
  time series can be achieved by computing the phase speed 
  (in the observer's frame) from the ratio of the electric field 
  to the magnetic field and using the relation $v_\mathrm{ph} = \omega/k$
  along the stream. This method is, however, applicable when 
  the electromagnetic component of the electric field is used 
  and when the electric field is not superposed or mixed over multiple waves.
  Generally, in the inertial range of solar wind turbulence 
  the fluctuation phase speeds are low and the Taylor hypothesis works well. 
  It does not work where its assumptions are violated
  The example of the foreshock and magnetosheath is intriguing 
  in that regard, but one has to be very careful in the
  foreshock where the entire concept of plasma moments is questionable due to the 
  non-Maxwellian nature of the solar wind distribution functions. 
  The magnetosheath might be a fruitful area to consider. 
}

In summary we conclude that the Taylor hypothesis can be safely applied under rather weak assumptions to the turbulent power spectrum of the velocity in both cases of advection by large eddies and in moderately to fast streaming plasmas with the restriction that one must take into account the relative directions of the mean stream velocity and the directions of the wavenumber of the turbulent velocities (eddies). In these cases the internal dispersion relation of the turbulence can be ignored if only the advection or streaming speeds are large enough. The wavenumber spectrum is then conserved in its either K or IK shape and Taylor-Galilei transforms with same shape into the spacecraft frequency frame. 

The same conclusion does, however, not rigorously hold for the magnetic power spectral density. This we have demonstrated in the main text and in full rigour in the  Appendices. Even though the amplitude of the magnetic field itself is not vulnerable to the transformation, the full spectrum of the turbulent fluctuations of the velocity appears in the exponential of the Fourier transform of the magnetic field principally inhibiting interpretation via the Taylor hypothesis. Application of Taylor's hypothesis to the magnetic power spectra then requires a complete neglect of any relation between the turbulent  magnetic and velocity fields. This is equivalent to the assumption that the magnetic field is by itself turbulent, which is unphysical. Nevertheless, assuming that this holds approximately, it implies that the Taylors transformation can approximately be applied if and only if the energy contained in the velocity power spectrum is completely negligible compared to the streaming energy density $\frac{1}{2}m_iN_0V_0^2$, i.e. the observations take place in a high speed flow. Only in this case one can conclude from the frequency magnetic power spectral density as measured in the spacecraft frame on the wavenumber spectrum of the turbulence.

\section*{Appendix A}
We  performed some of the integrations in Eq. (\ref{eq-final}) under the assumption that only one single turbulent wave mode $(w_0,\vec{\kappa}_0)$ dominates the turbulent velocity spectrum $\delta\vec{V}(t,\vec{x})$. 

Here, we will be more general because any  mode of frequency $w_0$ might be highly degenerated in the sense that it consists of a broad spectrum of modes with same frequency but completely different wavenumbers $\vec{\kappa}$ as, for instance, is the case in turbulent nonlinear sideband generation. Hence one must allow for an undefined broad spectrum in wave numbers $\vec{\kappa}$ even when taking only one frequency $w=w_0$. This implies that the general case concerning $\vec{\kappa}$ cannot be avoided. Then we have
\begin{equation}
\delta\vec{V}(t,\vec{x})=\frac{1}{16\pi^4}\int dw\,d\vec{\kappa}\:\delta\vec{V}_{w\vec{\kappa}}\Big\{\cos\Big[(w + \vec{\kappa\cdot V}_0) t+\vec{\kappa\cdot x}\Big]-i\sin\Big[(w + \vec{\kappa\cdot V}_0) t-\vec{\kappa\cdot x}\Big]\Big\}
\end{equation}
This results in a complicated time and space dependence of the magnetic fluctuations
\begin{equation}
\delta\vec{B}(t,\vec{x})=\frac{1}{16\pi^4}\int d\omega\,d\vec{k}\:\delta\vec{B}_{\omega\vec{k}} e^{-i\big[\omega(\vec{k}) + \vec{k\cdot }\big(\vec{V}_0+\delta\vec{V}(t,\vec{x})\big)\big]t+i\vec{k\cdot x}}
\end{equation}
One is interested in the spectrum of the fluctuations in the spacecraft frame, not the fluctuations themselves. The Fourier spectrum of velocity fluctuations in the approximation that just the main streaming velocity appears in its Fourier components is of course trivially obtained as the Fourier inversion of the velocity fluctuation spectrum (though, one could also retain the fluctuations themselves in the exponent and successively expand the integral under the assumption that the expansion converges). For the magnetic fluctuations, however, we retain the velocity fluctuations in the exponential and integrate with respect to time in Eq. (\ref{eq-final}) leaving the spatial integration for later. 

To make this explicit we define the integral operator  
\begin{equation}\label{ok}
\mathcal{O}_{\vec{k}}\equiv \frac{\vec{k}}{16\pi^4}\vec{\cdot}\int dw\,d\vec{\kappa}\: \delta\vec{ V}_{w\kappa}
\end{equation}
which when operating on $\exp\vec{\kappa\cdot x}$ depends on space $\vec{x}$ but in addition acts to the right on all functions containing $w$ and $\kappa$. Then we have for the magnetic fluctuations
\begin{eqnarray}\label{bfluc}
\delta\vec{B}_{\varpi\vec{K}}&=&\frac{1}{16\pi^4}\int d\omega\,d\vec{k}\:\delta\vec{B}_{\omega\vec{k}}\int dt\:d\vec{x}\:e^{-i[(\omega+\vec{k\cdot V}_0-\varpi)t+\vec{(K-k)\cdot x}]}\nonumber\\[-2ex]
&&\\[-2ex]
&&\qquad\times
e^{-it\mathcal{O}_{\vec{k}}\{\cos[(w+\vec{\kappa\cdot V}_0)t+\vec{\kappa\cdot x}]-i\sin[(w+\vec{\kappa\cdot V}_0)t-\vec{\kappa\cdot x}]\} }\bigg\}\nonumber
\end{eqnarray}
The time and space integrations have become separated. They  can in principle be independently performed. It is, however, more convenient to define
$\alpha\equiv \omega-\varpi+\vec{k\cdot V}_0$, and $\beta\equiv w+\vec{\kappa\cdot V}_0$. Integration is a complicated procedure. The presence of the trigonometric functions in the exponent indicates a strong nonlinear coupling. The sinus-term is real and will to lowest order be proportional to $t^2$. It hence introduces some kind of dissipation. However, for all real $\alpha, \beta$ the arguments of the trigonometric functions vary between $0$ and $\pi$, and dissipation may be small even though it will be finite. 

The first order relevant contribution corresponding to the Taylor hypothesis is obtained when introducing the new variable $\tau=\beta t$, expanding the trigonometric functions and retaining only the lowest order in $\tau$. The implication is that $\delta(\beta)=\delta(w+\vec{\kappa\cdot V}_0)$ is applied, and therefore
\begin{equation}
w=-\vec{\kappa\cdot V}_0
\end{equation}
is to be used in the integral for $\mathcal{O}$. This then yields for the  time integral
\begin{equation}
\frac{1}{\beta}\int d\tau\:e^{-i\tau(\alpha+\mathcal{O})/\beta}=2\pi\delta(\alpha+\mathcal{O})
\end{equation}
The $w$-operator function is thus  lost, and $\mathcal{O}$ becomes the transform of $\delta\vec{V}_{\vec{w\kappa}}$:
\begin{equation}
\mathcal{O}_{\vec{k}}(\vec{x})= \frac{1}{8\pi^3}\int d\vec{\kappa}\:\vec{k\cdot}\delta\vec{V}_{\!\!\!-\vec{\kappa\cdot V}_0,\vec{\kappa}}\:e^{i\vec{\kappa\cdot x}} \equiv \vec{k\cdot}\delta\vec{\mathcal{V}}_\mathit{turb}(\vec{V}_0,\vec{x})
\end{equation}
Though by now this is just a function, it enters at a very complicated place. The $\delta$-function requires that the frequency in $\delta\vec{B}_{\omega\vec{k}}$ is to be replaced by 
\begin{equation}
\omega=\varpi-\vec{k\cdot}[\vec{V}_0-\delta\vec{\mathcal{V}}_\mathit{turb}(\vec{V}_0,\vec{x})]
\end{equation}
before the integration with respect to $\vec{x}$ is performed. Here $\delta\vec{\mathcal{V}}_\mathit{turb}$ is the real space velocity fluctuation. This is the frequency $\varpi$ in the observer's frame shifted by both the streaming and some contribution of the turbulence, which however is still to be integrated over space before yielding the magnetic fluctuation amplitude. It shows that the Taylor hypothesis has a much more complicated consequence that the shift in frequency implies. 

Actually, this last integration cannot be easily performed even in this simplest case where we took only the lowest order approximation in the time integral. Because plugging in the expression for $\omega$, the Fourier transformed of the magnetic fluctuation amplitude obeys the following implicit representation:
\begin{equation}
\delta\vec{B}_{\varpi\vec{K}}=\frac{1}{8\pi^3}\int d\vec{k}\int d\vec{x}\:\delta\vec{B}_{\varpi-\vec{k\cdot}[\vec{V}_0-\delta\vec{\mathcal{V}}_\mathit{turb}(\vec{V}_0,\vec{x})],\vec{k}}\:e^{-i(\vec{k-K})\vec{\cdot x}}
\end{equation}
Here the spatial dependence is in the index on the magnetic fluctuation amplitude, which cannot be further resolved unless by iteration. Thus even the most crude approximation which takes into account the contribution of the velocity fluctuations in the Taylor transformation leads to a rather complicated dependence of the frequency on the spectrum of fluctuations in the mechanical turbulence. Resolution of the equation for the Fourier amplitude of the magnetic fluctuations becomes a formidable task and can be done only if completely neglecting the effect of the mechanical turbulence. This is what the Taylor hypothesis in fact imposes, and it is justified only when the flow speed by far exceeds the contribution of the fastest speeds in the fluctuations. The condition under which this is satisfied is high flow speeds $V_0\gg V_A$, far above the Alfv\'en speed $V_A$, which usually requires also that the ambient magnetic field is weak. 

\section*{Appendix B}
{The integral in Eq. (\ref{eq-final})
 can be expanded into Bessel functions when referring to the identities}
\begin{equation}
e^{-z\sin\tau}=\sum_n\,i^nI_n(z)e^{-in\tau}, \qquad e^{-iz\cos\tau}=\sum_mJ_m(z)e^{-im(\tau+\pi/2)}
\end{equation}
With these expressions Eq. (\ref{bfluc}) can be brought into the form
\begin{equation}
\delta\vec{B}_{\varpi\vec{K}}=\frac{1}{16\pi^4}\int d\omega\,d\vec{k}\:\delta\vec{B}_{\omega\vec{k}}\int dt\:d\vec{x}\sum_{nm}\:i^{n-m}e^{-i\alpha t-i[\vec{K-k}+(m-n)\vec{\kappa}]\vec{\cdot x}}I_n(t\mathcal{O}_{\vec{k}})J_m(t\mathcal{O}_{\vec{k}})\,e^{-i(n+m)\beta t}
\end{equation}
where now the integral operator $\vec{\mathcal{O}_{k}}$ Eq. (\ref{ok}) does not anymore depend on $\vec{x}$ as the spatial dependence has been absorbed into the Bessel expansion and appears only in the exponent of the exponential. Therefore this time the $\vec{x}$-integration can immediately be done. It yields the factor $8\pi^3\delta\big[\vec{K-k}+(m-n)\vec{\kappa}\big]$. In order to exploit the orthogonality of Bessel functions we transform $I_n(z)=i^{-n}J_n(iz)$. Orthogonality then requires that $n=m$, and we obtain
\begin{equation}
\delta\vec{B}_{\varpi\vec{K}}=\frac{1}{2\pi}\int d\omega\,d\vec{k}\:\delta\vec{B}_{\omega\vec{k}}\delta(\vec{K-k})\int dt\:\sum_{n}\:i^{-n}J_n(it\mathcal{O}_{\vec{k}})J_n(t\mathcal{O}_{\vec{k}})\,e^{-i(\alpha+2n\beta) t}
\end{equation}
The $t$-integration can also be done. It  includes only the product of the Bessel functions and the exponential, giving 
\begin{equation}
\int dt\:e^{-i(\alpha+2n\beta)t}J_n(it\mathcal{O}_{\vec{k}})J_n(t\mathcal{O}_{\vec{k}})=\frac{e^{-i\frac{\pi}{4}}}{\pi\mathcal{O}_{\vec{k}}}\,Q_{n-\frac{1}{2}}\Big[\frac{(\alpha+2n\beta)^2}{2i\mathcal{O}_{\vec{k}}^2}\Big]
\end{equation}
where $Q_\nu(z)$ is Legendre's function of the second kind. Then the final result becomes
\begin{equation}\label{b-final}
\delta\vec{B}_{\varpi\vec{K}}=\frac{1}{2\pi}\int d\omega\:\delta\vec{B}_{\omega\vec{K}}\sum_n\frac{i^{-n-\frac{1}{2}}}{\pi}\,\mathcal{O}_{\vec{K}}^{-1}Q_{n-\frac{1}{2}}\bigg\{\frac{1}{2i}\mathcal{O}_{\vec{K}}^{-2}\big[\omega-\varpi+\vec{K\cdot V}_0+2n(w+\vec{\kappa\cdot V}_0)\big]^2\bigg\}
\end{equation}
In this way one has obtained an integral equation for the Fourier amplitude of the magnetic fluctuations. Again, this cannot be treated further. It must be solved iteratively. On the other hand, this general expression has to be used in the power spectral density of the magnetic field, which leads to further difficulties. 

Even though the analytical result in the above form is intriguing, the dependence on the operator function $\mathcal{O}_{\vec{K}}$ in the infinite sum is rather complicated. This operator acts on $w$ and $\vec{\kappa}$. It  appears in its inverse in the pre-factor and also in the argument of the Legendre function where it is  raised to the second power. This makes the interpretation of this solution quite subtle. The required resolution of the $\omega$ integral needed for the isolation of the magnetic fluctuation amplitude becomes practically inhibited by this. 

We may, however, to lowest approximation assume that $w=-\vec{\kappa\cdot (V}_0-\vec{c}_s)$, where $\vec{c}_s$ is the inertial range eddy velocity which, for Kolmogorov scaling, becomes
\begin{equation}
c_s \sim (\ell \epsilon)^\frac{1}{3} \sim  \epsilon^\frac{1}{3}\kappa^{-\frac{1}{3}}
\end{equation}
Eddies in this case are an equivalent to some ``turbulent sound'', i.e. they are of linear dispersion. We do not argue whether or not this is true, which it is probably not, because there is no obvious reason that any turbulent motion resembles sound of linear dispersion. But it might hold for the largest mechanical turbulent eddies, at least as an approximation. In general $\vec{c}_s(w,\vec{\kappa})$ will depend on the eddy frequency and wavenumber and thus giving the mechanical equivalent of a ``turbulent dispersion relation''. However, if the streaming speed is very large by far exceeding the eddy speed, then the latter can be ignored and we have $w=-\vec{\kappa\cdot V}_0$, which is a  strongly simplifying though violent assumption because it assumes that the mechanical turbulence already obeys Taylor's assumption. On the other hand, this is not completely wrong. The mechanical turbulence for eddy velocities well below the velocity of light (which may be true in a wide range of applications) is indeed Galilei invariant. Therefore transport of the turbulent eddies by a very fast flow is not completely unreasonable. Then again $\mathcal{O}_{\vec{K}}\equiv \vec{K\cdot}\delta\vec{\mathcal{V}}^\mathit{turb}_{\vec{K\cdot V}_0}$ looses its operator function, and one has
\begin{equation}\label{taytay}
\delta\vec{B}_{\varpi\vec{K}}=\frac{1}{2\pi}\int d\omega\:\delta\vec{B}_{\omega\vec{K}}\sum_n\frac{i^{-n-\frac{1}{2}}}{\pi\vec{K\cdot}\delta\vec{\mathcal{V}}^\mathit{turb}_{\vec{K\cdot V}_0}}\,Q_{n-\frac{1}{2}}\bigg\{\frac{1}{2i}\bigg[\frac{\omega-\varpi+\vec{K\cdot V}_0}{\vec{K\cdot}\delta\vec{\mathcal{V}}^\mathit{turb}_{\vec{K\cdot V}_0}}\bigg]^2\bigg\}
\end{equation}
which is a much simpler relation than the former expression for the magnetic fluctuations. Still, the integral cannot be evaluated to isolate the spectral transform of the  magnetic fluctuation amplitude. This would require performing the $\omega$ integration, which can formally only be done if the $Q$-function could be replaced by a $\delta$-function, i.e. it should be strongly peaked around $\varpi=\omega+\vec{K\cdot V}_0$, again corresponding to a somewhat modified  Taylor assumption, but now already for the spacecraft frequency $\varpi$. This is unsatisfactory, however, because it reduces one to approach the initial assumption of its unrestricted validity.

Nevertheless doing this, one replaces $Q_{n-1/2}(z)\to 2\pi\delta(z)$. We then have
\begin{equation}
\delta\vec{B}_{\varpi\vec{K}}=\sum_n\frac{i^{-n+\frac{1}{2}}}{4\pi^4}\int d\omega\:\delta\vec{B}_{\omega\vec{K}}dw\,d\vec{\kappa}\vec{K\cdot}\delta\vec{V}_{w\vec{\kappa}}\delta\Big[\omega-\varpi+\vec{K\cdot V}_0+2n(w+\vec{\kappa\cdot V}_0)\Big]
\end{equation}
The argument of the $\delta$-function couples all the frequencies and wavenumbers. Performing the integration with respect to $\omega$ yields
\begin{equation}
\delta\vec{B}_{\varpi\vec{K}}=\sum_n\frac{i^{-n+\frac{1}{2}}}{4\pi^4}\int dw\,d\vec{\kappa}\,\delta\vec{B}_{\vec{K}[\varpi-\vec{K\cdot V}_0-2n(w+\vec{\kappa\cdot V}_0)]}\vec{K\cdot}\delta\vec{V}_{w\vec{\kappa}}
\end{equation}
This is the generalised Taylor assumption including already some severe assumptions and simplifications. The simplified Taylor hypothesis is recovered from this expression only if  in Eq. (\ref{taytay}) in addition to replacing $Q_{n-1/2}(z)\to 2\pi\delta(z)$ one restricts the sum to the zero order term $n=0$ which is achieved by multiplication with $2\pi\delta(n)$. In this case the sum degenerates, and one has 
\begin{equation}
\frac{2 i^{-\frac{1}{2}}}{\vec{K\cdot}\delta\mathcal{V}^\mathit{turb}_{\vec{K\cdot V}_0}}\delta\bigg[\bigg(\frac{\omega-\varpi+\vec{K\cdot V}_0}{\sqrt{2i}\vec{K\cdot\mathcal{V}}^\mathit{turb}_{\vec{K\cdot V}_0}}\bigg)^2\bigg]\Longrightarrow 2\sqrt{2}\:\delta\big(\omega-\varpi+\vec{K\cdot V}_0\big)
\end{equation}
where use has been made of the properties of the $\delta$-function. Inserting this into (\ref{taytay}) yields
\begin{equation}
\delta\vec{B}_{\varpi\vec{K}}\Longrightarrow\frac{\sqrt{2}}{\pi}\:\delta\vec{B}_{\varpi-\vec{K\cdot V}_0,\vec{K}}
\end{equation}
which, up to an unimportant numerical factor, is Taylor's suggestion if one sets the turbulent frequency $\varpi\ll\vec{K\cdot V}_0$ which translates trivially and effortlessly into the magnetic power spectral density.  It, however, implies complete neglect of the connection between the temporal and spatial spectra of both velocity (mechanical) and magnetic fluctuations, viz. any ``turbulent dispersion relation''.  

In order to be as precise as possible, one must return to Eq. (\ref{b-final}) and apply it iteratively, assuming that the naive Taylor hypothesis yields the zeroth-order approximation for $\delta\vec{B}_{\omega\vec{K}}$ and iterating the integral until convergence is achieved. This still implies some assumption about $\delta\vec{\mathcal{V}}$. Hence a model of the mechanical turbulence must in addition be imposed. Still, this procedure just produces the magnetic fluctuation spectrum which is then subject to the calculation of the magnetic power spectral density as done above for a particular case, which by itself is not simple.
 
We  thus conclude from this relatively rigorous treatment that the correct inclusion of the effect of the velocity turbulence into the use of the Taylor hypothesis in magnetohydrodynamic turbulent flows leads to difficulties in relating the Fourier amplitudes of the magnetic fluctuations measured in the observer's (spacecraft) frame to the wavenumber spectrum of the turbulence. It's application is restricted to high speed flows only where the turbulent velocities are vanishingly small against the stream \emph{on all scales} and without any exception including large-scale eddies. Though this may be a reasonably rough simplification, concerning the understanding of real magnetohydrodynamic turbulence in a flow it turns out to be rather unsatisfactory as it  barely allows for any reliable conclusions about scales, inertial and dissipation ranges, and the physics behind, unless the bulk streaming speed beats every other turbulent velocity by lengths.







\subsection*{Acknowledgments}
This work was part of a brief Visiting Scientist Programme granted at the International Space Science Institute Bern. We acknowledge the interest of the ISSI directorate as well as the generous hospitality of the ISSI staff. 



\end{document}